\documentclass{llncs}
\usepackage{pst-node}
\usepackage{latexsym}
\usepackage{amsmath}
\usepackage{epsfig}
\usepackage{makeidx}  

\def\q5uad{\quad\quad\quad\quad\quad}

\begin{document}
\mainmatter              

\title{Watermarking PDF Documents using Various Representations of Self-inverting Permutations}

\titlerunning{Watermarking PDF Documents}  

\author{Maria~Chroni \ and \ Stavros~D.~Nikolopoulos}
\authorrunning{Chroni and Nikolopoulos} 
%
\tocauthor{Maria~Chroni and Stavros~D.~Nikolopoulos}
\institute{Department of Computer Science \& Engineering, \\
University of Ioannina\\
GR-45110 Ioannina, Greece\\
\email{\{mchroni,stavros\}@cs.uoi.gr}}

\maketitle              

\begin{abstract}
This work provides to web users copyright protection of their Portable Document Format (PDF) documents by proposing efficient and easily implementable techniques for PDF watermarking; our techniques are based on the ideas of our recently proposed watermarking techniques for software, image, and audio, expanding thus the digital objects that can be efficiently watermarked through the use of self-inverting permutations. In particular, we present various representations of a self-inverting permutation $\pi^*$ namely 1D-representation, 2D-representation, and RPG-representation, and show that theses representations can be efficiently applied to PDF watermarking. Indeed, we first present an audio-based technique for marking a PDF document $T$ by exploiting the 1D-representation of a permutation $\pi^*$, and then, since pages of a PDF document $T$ are 2D objects, we present an image-based algorithm for encoding $\pi^*$ into $T$ by first mapping the elements of $\pi^*$ into a matrix $A^*$ and then using the information stored in $A^*$ to mark invisibly specific areas of PDF document $T$. Finally, we describe a graph-based watermarking algorithm for embedding a self-inverting permutation $\pi^*$ into the document structure of a PDF file $T$ by exploiting the RPG-representation of $\pi^*$ and the structure of a PDF document. We have evaluated the embedding and extracting algorithms by testing them on various and different in characteristics PDF documents.
\end{abstract}


\section{Introduction}
\label{sec:introduction}
\noindent Information age has altered the way people communicate by breaking the barriers imposed on communications by time, distance, and location and has undoubtedly impact not only humans activities but also global industry and economy. Communication has been greatly affected by the constant and rapid evolution of many technologies such as fiber optic, cellular and satellite technology, networking, digital transmission and compression as well as advanced computers, and improved human-computer interaction. The aforementioned technologies allow the rapid transmission, and store, of great amounts of information.

The digital era has already had extensive impacts on business, commerce, education, services, and social life. The concepts of e-government, e-learning, e-commerce, e-business, e-publishing, refer peoples' interaction in the digital world. In this world, people everyday, interact by exchanging e-mails, instant messages, video, audio, images, and digital documents. Part of the information transmitted is an increasing amount of sensitive information, such as personal data, medical and financial records, business information, government data, legal documents. Another part of information available in the web is used to promote ones' work or product.

Electronic document, is an extensively used medium traveling over the internet for information exchange and due to the ease of copying and distributing they are susceptible to threats like illegal copying, redistribution of copyrighted documents, and plagiarism. Subsequently, it has become more important to protect the electronic documents from any malicious user while existing in the digital world. Copyright protection of digital contents is such a need of time which cannot be overlooked. In past, various methods like encryption, steganography and watermarking has been used to solve these problems. However, digital watermarking is the better solution for copyright protection than encryption and steganography. It is well known that digital watermarking methods are efficient enough to identify the original copyright owner of the contents.

Recall that there are many reasons why you would want to use watermarks in digital documents: as a copying deterrent, as a means of identifying the source of a printed document, as a means of determining whether a document has been altered, etc.

\vspace*{0.2in}
\noindent {\bf Attacks.} Any action that a user can perform on a text that can affect the watermark, or its usefulness, is called attack. In \cite{Zhou2009} existing attacks on text watermarking can be classified into three main categories:

\begin{itemize}
\item[$\circ$\,] {watermark attacks},
%
\item[$\circ$\,] {geometric attacks}, and
%
\item[$\circ$\,] {system attacks}.
\end{itemize}

\noindent In a watermark attack, the adversary aims to detect and destroy the watermark, without necessarily decoding the original message. In contrast to watermark attacks, geometrical attacks are blind attacks on watermarked text documents. The process of these attacks requires neither the algorithmic knowledge of the watermarking technique nor the watermarking key, geometrical attacks intend not to remove the embedded watermark itself, but to prevent it from serving its intended purpose through altering format or content of the watermarked text documents. This type of attack includes reformatting, reproducing, sentences swapping, paragraphs shuffling, the addition/deletion of words, sentences and paragraphs. System attacks use several signal processing tools such as principal component analysis, independent component analysis, clustering, vector quantization, etc.

\vspace*{0.2in}
\noindent {\bf Related Work.}~Text watermarking is the area of research that has emerged after the development of Internet and communication technologies; we mention that the first reported effort on marking documents dates back to 1993~\cite{Maxemchuk1997}.

Generally, we can classify the previous work on digital text watermarking in the following four categories:

\begin{itemize}
\item[$\circ$\,] {image based approach},
%
\item[$\circ$\,] {syntactic approach},
%
\item[$\circ$\,] {semantic approach}, and
%
\item[$\circ$\,] {structural approach}.
\end{itemize}

\noindent In image-based approach, a watermark is embedded in text image. Brassil, et~al. were the first to propose a few text watermarking methods utilizing text image~\cite{Brassil1995a,Brassil1995b}; they also developed document watermarking schemes based on line shifts, word shifts as well as slight modifications to the characters \cite{Brassil1999}. Maxemchuk, et~al.~\cite{Maxemchuk1997,Maxemchuk1998,Maxemchuk1994} analyzed the performance of these methods, while later Low, et~al. \cite{Low1998,Low2000} further analyzed their efficiency. Huang and Yan \cite{Huang2001} proposed a text watermarking method based on an average inter-word distance in each line.

In syntactic approach, the syntactic structure of the text is used to embed watermark. Atallah, et~al.~\cite{Atallah2002} proposed several methods of natural language watermarking, which opened up a brand-new and challenging research direction for text watermarking. Meral et~al. performed morpho-syntactic alterations to the text to watermark it~\cite{Meral2009}; they also provided an overview of available syntactic tools for text watermarking~\cite{Meral2007}.

In semantic approach, semantics of text are used to embed the watermark in text. Atallah et~al. were the first to propose the semantic watermarking schemes~\cite{Atallah2002}. Later, the synonym substitution method was proposed, in which watermark was embedded by replacing certain words with their synonyms \cite{Topkara2006}. Sun, et~al. \cite{Sun2005} proposed noun-verb based technique for text watermarking which used nouns and verbs parsed by semantic networks. Topkara, et~al. proposed an algorithm of the text watermarking by using typos, acronyms and abbreviation in the text to embed the watermark~\cite{Topkara2007}. Algorithms were developed to watermark the text using the linguistic approach of presuppositions~\cite{Macq2007} in which the discourse structure, meaning, and representations are observed and utilized to embed watermark bits. The text pruning and the grafting algorithms were also developed in the past. Another algorithm based on text meaning representation (TMR) strings has also been proposed~\cite{Lu2008}.

The structural approach is the most recent approach used for copyright protection of text documents. In this approach, text is not altered, rather it is used to logically embed watermark in it. A text watermarking algorithm, for copyright protection of text using occurrences of double letters (aa-zz) in text, has recently been proposed \cite{Jalil2010}. Recently, a significant number of techniques have been proposed in the literature which use Portable Document Format (PDF) files as cover media in order to hide data~\cite{Bindra2011a,Bindra2011b,Liu2012,Liu2008,Liu2006,Lee2010,Zhong2007}.

\vspace*{0.2in}
\noindent {\bf Contribution.}~In this paper, in order to provide to web users copyright protection of their digital documents, we present easily implemented techniques for watermarking PDF documents. Our aim is to extent the digital objects that the proposed representations of a self-inverting permutation, i.e. the 1D-representation, the 2D-representation, and the RPG-representation, can be efficiently applied to; note that, RPG-representation means the encoding of permutation $\pi^*$ as a reducible permutation graph $F[\pi^*]$.

We first propose an image-based technique for marking the PDF document $T$ by exploiting the 1D-representation of a permutation $\pi^*$. The embedding of a mark is performed by increasing the distance (or, space) between two consecutive words in a paragraph of the document $T$. The extraction algorithm operates in a reverse manner.

Consequently, since pages of a PDF documents $T$ are two dimensional objects, we propose an algorithm for encoding a self-inverting permutation $\pi^*$ into a document $T$ by first mapping the elements of $\pi^*$ into an $n^* \times n^*$ matrix $A^*$ and then using the information stored in $A^*$ to mark invisibly specific areas of PDF document $T$ resulting thus the watermarked PDF document $T_w$. We also propose an efficient algorithm for extracting the embedded self-inverting permutation $\pi^*$ from the watermarked PDF document $T_w$ by locating the positions of the marks in $T_w$; it enables us to recontract the 2D representation of the self-inverting permutation $\pi^*$.

Finally, we describe a watermarking algorithm for embedding a self-inverting permutation into the document structure of a PDF file $T$, by exploiting the graph representation of $\pi^*$ and the structure of a PDF document $T$.
More precisely, in light of the two embedding algorithms ${\tt Encode\_SiP.to.RPG}$-${\tt I}$ and -${\tt II}$, we present an algorithm for embedding a reducible permutation graph $F[\pi^*]$ into a PDF document $T$. The main idea behind the proposed embedding algorithm is a systematic addition of appropriate object-references in the input PDF document $T$, through the use of entries of type ${\tt \backslash kye(\cdot)}$, so that the graph $F[\pi^*]$ can be easily constructed from the page tree PT($T_w$) of the resulting watermarked document $T_w$.

\vspace*{0.2in}
\noindent {\bf Road Map.} The paper is organized as follows: In Section~\ref{sec:Background-Results-PDF} we establish the notation and related terminology, and we present background results. In Section~\ref{sec:Watermarking-PDF-Document} based on the three different representations of self-inverting permutation (SiP), i.e., the 1D-representation, the 2D-representation, and the RPG-representation (the encoding of permutation $\pi^*$ as a reducible permutation graph $F[\pi^*]$), we present the algorithms ${\tt Embed\_SiP.to.PDF}$-${\tt I}$, ${\tt Embed\_SiP.to.PDF}$-${\tt II}$, and ${\tt Embed\_RPG.to.PDF}$, along with the corresponding extracting algorithms, for embedding a watermark number (or, equivalently, a self-inverting permutation $\pi^*$ or a reducible permutation graph $F[\pi^*]$) into a PDF document file. Finally, in Section~\ref{sec:Concluding-Remarks-PDF} we conclude the paper and discuss possible future extensions.

\section{Background Results}
\label{sec:Background-Results-PDF}
\noindent In this section we give some definitions and the theoretical background we use towards the watermarking of Portable Document Format (PDF) documents. We first briefly present the different representations of a self-inverting permutation (SiP), and then we present the structure of PDF documents.

\vspace*{0.2in}
\noindent \textbf{1D-representation of SiP.} Recently, we presented the one-dimensional representation (1D-representation) of a self-inverting permutation (SiP) $\pi^*$ and the one-dimensional marked representation of $\pi^*$ (1DM-representation), and showed how to embed a SiP, represented by 1D space, into an audio signal \cite{Chroni-PhD2014,CFN14}. In our 1D-representation, the elements of the permutation $\pi$ are mapped in specific cells of an array $B$ of size $n^2$ as follows:

\vspace{0.2cm}
%

\begin{itemize}
    \item[$\bullet$] number \,  $\pi_i$ \, $\longrightarrow$  \, entry \, $B((\pi^{-1}_{\pi_i}-1)n+\pi_i)$
\end{itemize}

\vspace{0.2cm}
\noindent or, equivalently, the cell at the position $(i-1)n+\pi_i$ is labeled by the number $\pi_i$, for each $i = 1, 2, \ldots, n$.

\vspace{0.2cm}
In our 1DM representation, a permutation $\pi$ over the set~$N_n$ is represented by an $n^2$ array $B^*$ by marking the  cell at the position $(i-1)n+\pi_i$ by a specific symbol, where, in our implementation, the used symbol is again the asterisk character ``*".

\begin{figure*}[t!]
    \hrule\medskip\medskip\smallskip
    \centering
    \includegraphics[scale=0.5]{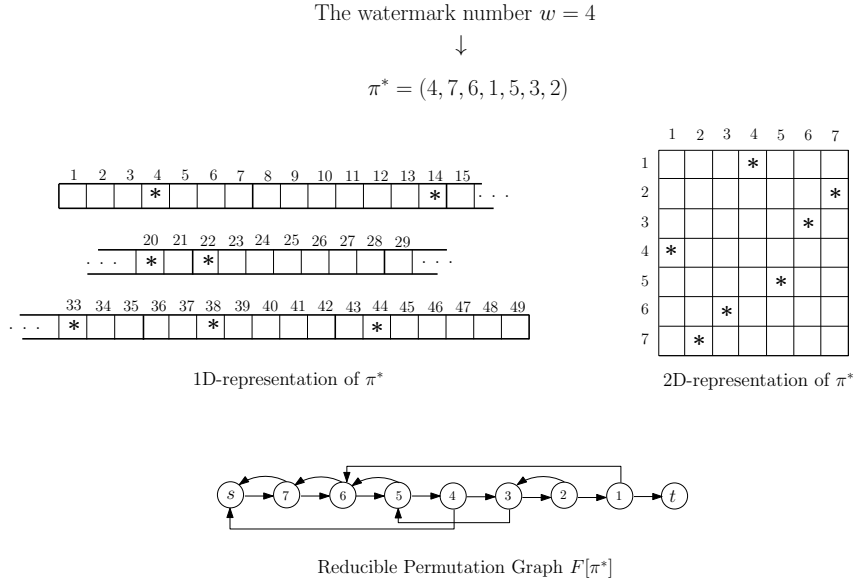}
    \centering
    \medskip\medskip\hrule\medskip\medskip\medskip
    \caption{\small{Three different representations of permutation $\pi^*=(4,7,6,1,5,3,2)$.}}
\medskip
\label{fig:Fig-1}
\end{figure*}

\vspace*{0.2in}
\noindent \textbf{2D-representation of SiP.} We have also presented the two-dimensional representation of a SiP (2D-representation) and the two-dimensional marked representation of SiP (2DM-representation); note that, theses representations have been recently used for watermarking images in the frequency domain \cite{Chroni-PhD2014,CFN14}.

We defined the 2D-representation of a SiP as the representation where the elements of the permutation $\pi=(\pi_1, \pi_2, \ldots ,\pi_n)$ are mapped in specific cells of an $n \times n$ matrix $A$ as follows:
\begin{itemize}
    \item[$\bullet$] number \,  $\pi_i$ \, $\longrightarrow$  \, entry \, $A(\pi^{-1}_i, \pi_i)$
\end{itemize}

\vspace*{0.06in}
\noindent or, equivalently,
\vspace*{0.06in}

\begin{itemize}
    \item[$\bullet$] the cell at row $i$ and column $\pi_i$ is labeled by the number $\pi_i$, for each $i = 1, 2, \ldots, n$.
\end{itemize}

\vspace{0.2cm}

\noindent In 2DM-representation the cell at row $i$ and column $\pi_i$ of matrix $A$ is marked by a specific symbol, for each $i = 1, 2, \ldots, n$.

We have presented algorithms for embedding the 2D-dimensional representation of SiP in an image. Recall that the matrix $A$ incorporates important structural properties which, in image watermarking, make it possible to detect geometric transformations on the watermarked image. The properties of the matrix $A$ are the following:
\begin{itemize}
  \item[$\circ$] the matrix $A$ is symmetric;
  \item[$\circ$] the main diagonal of the symmetric matrix $A^*$ has always one and only one marked cell;
  \item[$\circ$] the marked cell on the diagonal is always in entry $(i, i)$ of $A^*$, where $i = \lceil \frac{n^*}{2} \rceil + 1$, $\lceil \frac{n^*}{2} \rceil + 2$, $\dots$, $n^*$.
\end{itemize}

\noindent The authors of this paper, we have also presented an efficient and easily implemented algorithm for encoding numbers as reducible permutation graphs (or, for short, RPG) through the use of self-inverting permutations \cite{CN13,CN12}. In particular, we have proposed two such encoding algorithms: the algorithm ${\tt Encode\_SiP.to.RPG}$-${\tt I}$ applies to any permutation $\pi$ and relies on domination relations on the elements of $\pi$ whereas the algorithm ${\tt Encode\_SiP.to.RPG}$-${\tt II}$ applies to a self-inverting permutation $\pi^*$ produced in any way and relies on the decreasing subsequences of $\pi^*$. Figure~\ref{fig:Fig-1} summarizes by an example the representations of the permutation $\pi^* = (4,7,6,1,5,3,2)$.

\begin{figure*}[t!]
    \hrule\medskip\medskip\smallskip
    \centering
    \includegraphics[scale=0.38]{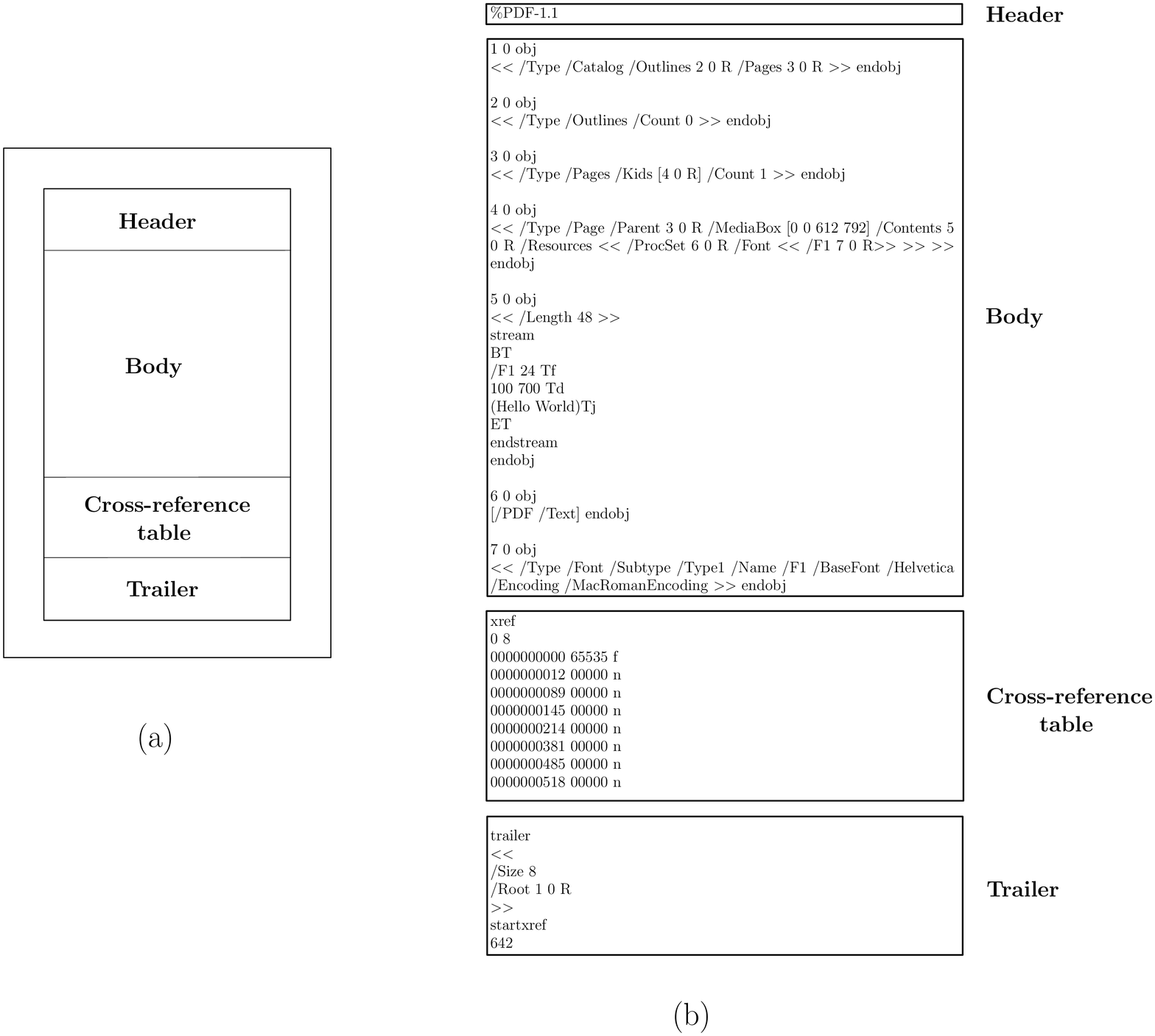}
    \centering
    \medskip\medskip\hrule\medskip\medskip\medskip
    \caption{\small{(a) The structure of a PDF file; (b) The code of a PDF file containing, in object ${\tt 5 \ 0 \ obj}$, the text ``Hello World".}}
\medskip
\label{fig:File-structure}
\end{figure*}

\vskip 0.0in
\subsection{Structure of a PDF Document}
\label{subsec:Structure-of-PDF}
\noindent The Portable Document Format (PDF)~\cite{Adobe2006} is an open standard (defined in ISO 32000) which facilitates device and platform independent capture and representation of rich information such as text, multimedia and graphics, into a single medium. Thus the PDF format enables viewing and printing of a rich document, independent of either application software or hardware. In this section we present a structural analysis of a PDF file, by giving its basic components.

\vspace{0.3cm}
\noindent {\bf Object.} An object is the basic element in PDF files, in which eight kinds of objects, namely Boolean Object, Numeric Object, String Object, Name Object, Array Object, Null Object, Dictionary and Stream Object are sustained. Objects may be labeled so that they can be referred to by other objects. A labeled object is called an indirect object.

\vspace{0.3cm}
\noindent {\bf File structure.} The PDF file structure determines how objects are stored in a PDF file, how they are accessed, and how they are updated. The file structure (see, Figure~\ref{fig:File-structure}) includes the following:

\vspace*{-0.05in}
\begin{itemize}
  \item[$\circ$] an one-line header identifying the version of the PDF specification to which the file conforms,
  \item[$\circ$] a body containing the objects that make up the document contained in the file,
  \item[$\circ$] a cross-reference table containing information about the indirect objects in the file, and
  \item[$\circ$] a trailer giving the location of the cross-reference table and of certain special objects within the body of the file.
\end{itemize}

\begin{figure*}[t!]
    \hrule\medskip\medskip\smallskip
    \centering
    \includegraphics[scale=0.35]{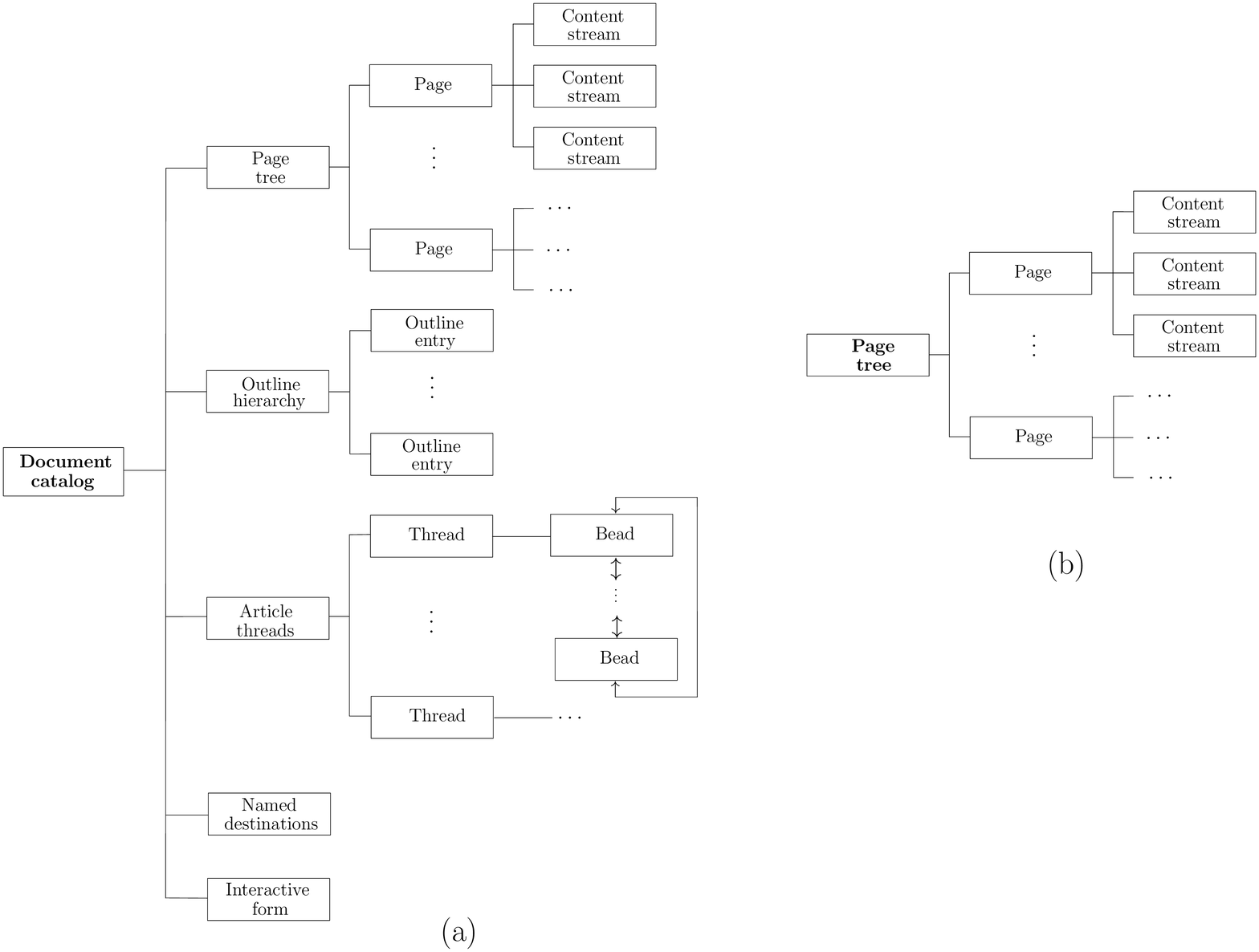}
    \centering
    \medskip\medskip\hrule\medskip\medskip\medskip
    \caption{\small{(a) The main structural components of a PDF file; (b) The document structure of PDF file.}}
\medskip
\label{fig:Document-structure}
\end{figure*}

\noindent Figure~\ref{fig:File-structure} shows an example of a PDF file and its internal file structure.

\vspace{0.3cm}
\noindent {\bf Document structure.} The PDF document structure specifies how the basic object types are used to represent components of a PDF document: pages, fonts, annotations, and so forth. The document structure of a PDF file is organized in the shape of an object tree topped by Catalog, Page tree, Outline hierarchy and Article thread included. The Outline hierarchy is the bookmarker of PDF, and Page tree includes page and Pages which in turn includes the total page number and each page marker. Page, the main body of PDF file, is the most important object which involves the typeface applied, the text, pictures, page size, and so on. The organization of other objects is analogous to Page tree. Figure~\ref{fig:Document-structure} illustrates the structure of the object hierarchy.

\section{Watermarking PDF Documents}
\label{sec:Watermarking-PDF-Document}
\noindent In this section we describe embedding algorithms for encoding a SiP $\pi^*$ into a digital document $T$. More specifically, we embed the permutation $\pi^*$ into a PDF document by exploiting (i) the one-dimensional representation of $\pi^*$, (ii) the two-dimensional representation of a $\pi^*$, and (iii) the encoding of $\pi^*$ as a reducible permutation graph $F^*[\pi^*]$.

\vspace*{0.0in}
\subsection{Embed Watermark into PDF - I}
\label{subsec:Algorithm-Embed-SiP-to-PDF-I}
We first design an embedding algorithm for watermarking a PDF document by exploiting the 1D-representation of a permutation $\pi^*$. The marking is performed by increasing the space between two consecutive words in a paragraph of $T$.

Let $B^*$ be the 1D array of size $n = n^* \times n^*$ which represents the permutation $\pi^*$ of length $n^*$, and let (w$_1$, s$_1$), (w$_2$, s$_2$), $\ldots$, (w$_n$, s$_n$) be $n$ pairs of type ``word-space" of a paragraph {\it par} of the input PDF document; recall that the entry $B^*((i-1)n^* + \pi^*_i)$ contains the symbol ``*", $1 \leq i \leq n^*$. The algorithm increases by a small value ``c" the $i$-th space of the pair (w$_i$, s$_i$) if $B^*((i-1)n^* + \pi^*_i)=``*"$.

We next give a high-level description, with respect to PDF modification, of our proposed embedding algorithm.

\begin{figure*}[t!]
    \hrule\medskip\medskip\smallskip\smallskip
    \centering
    \includegraphics[scale=0.35]{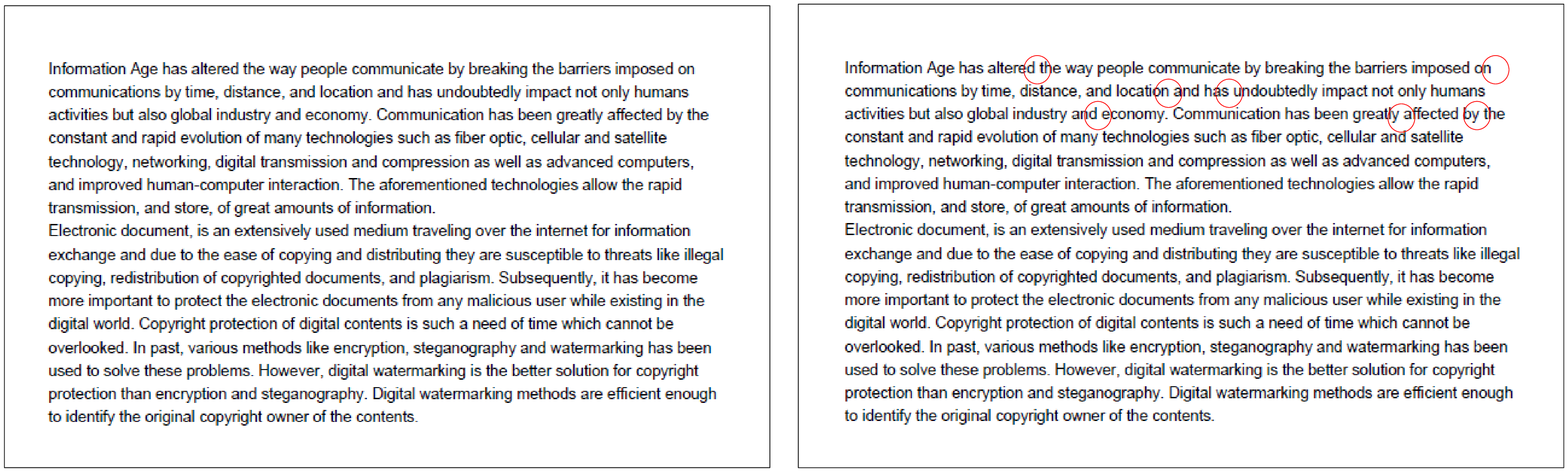}
    \centering\\
     {(a)} \hspace*{2.2in}  {(b)}\\
    \medskip\medskip\hrule\medskip\medskip\medskip
    \caption{\small{(a) The initial PDF document $T$; (b) The watermarked PDF document $T_w$ using the 1D-representation of permutation $\pi^* = (4,7,6,1,5,3,2)$; the red cycles indicate the marks.}}
\medskip
\label{fig:File-structure-ex}
\end{figure*}

\vspace*{0.2in}
\noindent {\bf Algorithm ${\tt Embed\_SiP.to.PDF}$-${\tt I}$}
\begin{enumerate}
\item[\textbf{1}.\,]   
Compute the 1DM representation of the permutation $\pi^*$, i.e., construct the array $B^*$ of size $n = n^* \times n^*$ where the $(i-1)n^* + \pi^*_i$ entry of $B^*$ contains the symbol ``*", $1 \leq i \leq n^*$;
\vspace*{0.05in}

\item[\textbf{2}.\,]   
Select an appropriate paragraph {\it par} on a page $P$ of PDF document $T$ to embed the self-inverting permutation $\pi^*$;
\vspace*{0.05in}

\item[\textbf{3}.\,]   
Partition the paragraph {\it par} into $n$ pairs $(w_1, s_1), (w_2, s_2), \ldots, (w_n, s_n)$, where $w_i$ and $s_i$ are the $i$-th word and space, respectively, in selected paragraph {\it par}, $1 \leq i \leq n$;
\vspace*{0.05in}

\item[\textbf{4}.\,]   
For each pair $(w_i, s_i)$ s.t. $B^*((i-1)n^* + \pi^*_i)=``*"$, increases the space $s_i$ or, equivalently, distance $d(w_i,w_{i+1})$ between words $w_i$ and $w_{i+1}$, by a relative small value $c$, $1 \leq i \leq n$;
\vspace*{0.05in}

\item[\textbf{5}.\,]   
Return the watermarked PDF document $T_w$.
\end{enumerate}

\vspace*{0.1in}
\noindent {\bf Extraction.}~The extraction algorithm, which we call ${\tt Extract\_PDF.from.SiP}$-${\tt I}$, operates as follow: it takes as input the watermarked PDF document $T_w$, locates the paragraph {\it par}, and computes the permutation $\pi^*$ by finding the positions of the words $w_i$ such that:

\begin{itemize}
\item[$\circ$ ] $d(w_i,w_{i+1}) > d(w_{i-1},w_{i})$, or

\item[$\circ$ ] $d(w_i,w_{i+1}) > d(w_{i+1},w_{i+2})$
\end{itemize}

\noindent where, $d(w_i,w_j)$ is the distance between words $w_i$ and $w_j$ in a paragraph {\it par} of $T_w$, $1 \leq i \leq n$; note that, an appropriate paragraph {\it par} contains more that $n$ words.

\vspace*{0.0in}
\subsection{Embed Watermark into PDF - II}
\label{subsec:Algorithm-Embed-SiP-to-PDF-II}
\noindent In this section we describe a different approach of embedding algorithm a self-inverting permutation $\pi^*$ into a digital document $T$, by exploiting the two-dimensional representation of permutation $\pi^*$.

The main idea behind the embedding algorithm, we call it ${\tt Embed\_SiP.to.PDF}$-${\tt II}$, is similar of that of algorithm ${\tt Embed\_SiP.to.Image}$-${\tt F}$ (see, \cite{CFN13}). The most important of this idea is the fact that it suggests a way in which the permutation $\pi^*$ can be represented with a 2D-representation and since pages of a PDF document $T$ are two dimensional objects that representation can be efficiently marked on them resulting the watermarked PDF document $T_w$; in a similar way as in our image watermarking approach, such a 2D-representation can be efficiently extracted for a watermarked PDF document $T_w$ and converted back to the self-inverting permutation $\pi^*$.

Let $A^*$ be the 2D matrix of size $n^* \times n^*$ which represents the permutation $\pi^*$ of length $n^*$. The marking of the input PDF document $T$ is performed by selecting an appropriate page $P$ of $T$ and setting $n^*$ objects (e.g., characters, symbols, images) in a specific positions on page $P$, $1 \leq i \leq n^*$. In fact, we set an object $\mathcal{O}_i$ in position with $(x'_i,y'_i)$ coordinates on page $P$ if $A^*(x_i,y_i)=``*"$, where $1 \leq x_i, y_i \leq n^*$ and $0 \leq x'_i, y'_i \leq size(P)$; note that, $(0,0)$ is the lower-left point (or, equivalently, the bottom-left corner) of the page $P$.

The algorithm takes as input a SiP $\pi^*$ and a PDF document $T$, and returns the watermarked document $T_w$; it consists of the following steps.

\begin{figure*}[t!]
    \hrule\medskip\medskip\smallskip
    \centering
    \includegraphics[scale=0.26]{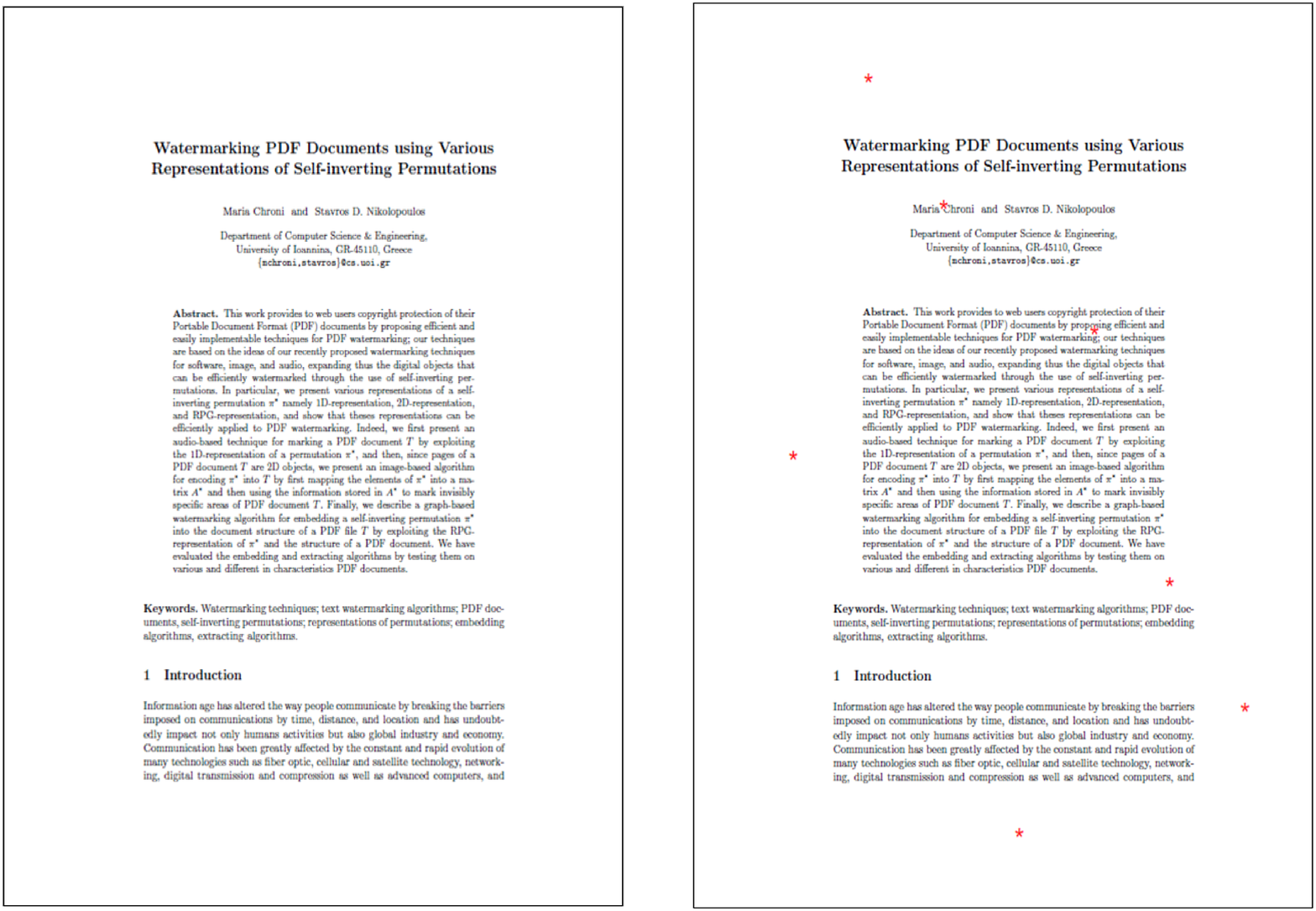}
    \centering\medskip\\
     {(a)} \hspace*{2.0in}  {(b)}\\
    \medskip\medskip\hrule\medskip\medskip\medskip
    \caption{\small{(a) The initial PDF document $T$; (b) The watermarked PDF document $T_w$ using the 2D representation of permutation $\pi^* = (4,7,6,1,5,3,2)$; the red stars indicate the marks.}}
\medskip
\label{fig:File-structure-ex}
\end{figure*}

\vspace*{0.2in}
\noindent {\bf Algorithm ${\tt Embed\_SiP.to.PDF}$-${\tt II}$}
\begin{enumerate}
\item[\textbf{1}.\,]   
Compute the 2DM representation of the self-inverting permutation $\pi^*$, i.e., construct an array $A^*$ of size $n^* \times n^*$ s.t. the entry $A^*(i,\pi^*_i)$ contains the symbol ``*", $1 \leq i \leq n^*$;
\vspace*{0.00in}

\item[\textbf{2}.\,]   
Select an appropriate page $P$ to embed the permutation $\pi^*$ and compute the size $size(P)$ of the page $P$, say, $N \times M$;
\vspace*{0.00in}

\item[\textbf{3}.\,]   
Segment the PDF page $P$ into $n^* \times n^*$ grid-cells $C_{ij}$ of size $\left \lfloor \frac{N}{n^*} \right \rfloor \times \left \lfloor \frac{M}{n^*} \right \rfloor$, $1 \leq i, j \leq n^*$;
\vspace*{0.00in}

\item[\textbf{4}.\,]   
For each grid-cell $C_{ij}$ s.t. $A^*(i,j)=``*"$, mark the cell $C_{ij}$ by setting a symbol, with an appropriate color, in any position inside $C_{ij}$ of $P$, $1 \leq i, j \leq n^*$, resulting thus the marked document $T_w$;
\vspace*{0.00in}

\item[\textbf{5}.\,]   
Return the watermarked PDF document $T_w$.

\end{enumerate}

\begin{figure*}[t!]
    \hrule\medskip\medskip\smallskip
    \centering
    \includegraphics[scale=0.38]{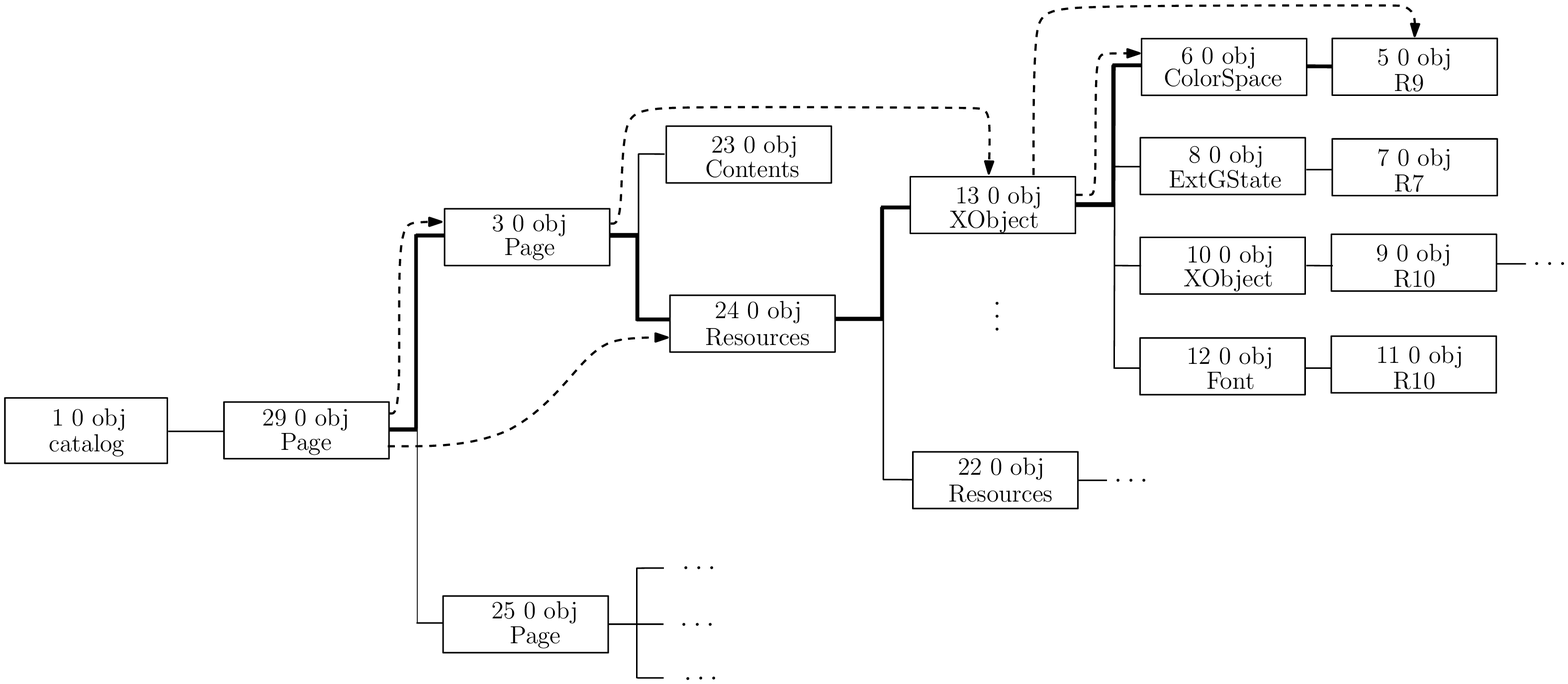}
    \centering
    \medskip\medskip\medskip\hrule\medskip\medskip\medskip
    \caption{\small{The watermarked $DS(T_w)$ which encodes the RPG of $\pi^* = (4,5,3,1,2)$.}}
\medskip
\label{fig:RPG-in-Page-tree}
\end{figure*}

\vspace*{0.1in}
\noindent {\bf Extraction.}~The algorithm which extracts the permutation $\pi^*$ from the watermarked PDF $T_w$ operates in a similar way as the corresponding extraction algorithm for images: it takes the input watermarked image $I_w$, locate the marked page $P$, computes its $N \times M$ size, and segments $P$ into $n^* \times n^*$ grid-cells $C_{ij}$ of size $ \left \lfloor \frac{N}{n^*} \right \rfloor \times \left \lfloor \frac{M}{n^*} \right \rfloor$; then, it computes the permutation $\pi^*$ by finding the coordinates $(x_i, y_i)$ of the $n^*$ symbols in the page $P$, $1 \leq i \leq n^*$; we call it ${\tt Extract\_PDF.from.SiP}$-${\tt II}$.

\vspace*{0.0in}
\subsection{Embed an RPG into a PDF}
\label{subsec:Algorithm-Embed-RPG-to-PDF}

\noindent In this section we describe a watermarking algorithm for embedding a self-inverting permutation $\pi^*$ into a PDF document $T$, by exploiting the RPG-representation of $\pi^*$ and the structure of a PDF document $T$.

Indeed, we have recently proposed two algorithms, namely ${\tt Encode\_SiP.to.RPG}$-${\tt I}$ and -${\tt II}$, for encoding self-inverting permutations $\pi^*$ as reducible permutation graphs $F[\pi^*]$. Moreover, in this paper we have described the document structure DS($T$) of a PDF document $T$ (see, Subsection~\ref{subsec:Structure-of-PDF}); note that, the document structure of a PDF file always contains a node, namely ${\tt Document}$-${\tt catalog}$, and a page tree PT($T$) rooted at node ${\tt Page}$-${\tt tree}$, denoted by ${\tt root(pt)}$; see, Figure~\ref{fig:Document-structure}(b).

In light of the two encoding algorithms ${\tt Encode\_SiP.to.RPG}$-${\tt I}$ and -${\tt II}$, we next present an algorithm for embedding a reducible permutation graph $F[\pi^*]$ into a PDF document $T$. The main idea behind the proposed embedding algorithm is a systematic addition of appropriate object-references in selected nodes of the page-tree PT($T$) of the document structure DS($T$), through the use of entries of type ${\tt /Kye(\cdot)}$, so that the graph $F[\pi^*]$ can be easily constructed from the page-tree PT($T_w$) of the resulting watermarked document $T_w$.

Let $F[\pi^*]$ be a reducible permutation graph produced by one of our two encoding algorithms (i.e., ${\tt Encode\_SiP.to.RPG}$-${\tt I}$ or -${\tt II}$), and let $u_{n+1}, u_{n}, \ldots, u_1, u_{0}$ be the nodes of the graph $F[\pi^*]$; note that, $F[\pi^*]$ does not contain the back-edge $(u_{0}, u_{n+1})$. In order to simplify the extraction process, the graph $F[\pi^*]$ which is embedded into a PDF document $T$ contains one extra back-edge, i.e., the edge $(u_{0}, u_{n+1})$; see, \cite{CN13,CN12}.

The algorithm for embedding a reducible permutation graph $F[\pi^*]$ into a PDF document $T$ is called ${\tt Encode\_RPG.to.PDF}$ and is described below.

\vskip 0.0in 
\vspace*{0.2in}
\noindent {\bf Algorithm ${\tt Encode\_RPG.to.PDF}$}
\begin{enumerate}
\item[\textbf{1}.\,]
Compute the document structure DS($T$) of the input PDF document $T$ and locate its page-tree PT($T$); let ${\tt node(dc)}$ be the document catalog node of structure DS($T$) and ${\tt root(pt)}$ be the root node of the page tree PT($T$); see, Figure~\ref{fig:Document-structure}(b);
\vspace*{0.05in}
\item[\textbf{2}.\,]
Compute a path $O(T)=(v_{n+1}, v_{n}, \ldots, v_1, v_{0})$ on $n+2$ nodes (i.e., objects) of the page-tree PT($T$) s.t. $v_{n+1}=root(pt)$, and set $s=v_{n+1}$ and $t=v_{0}$;
\vspace*{0.05in}
\item[\textbf{3}.\,]
Assign an exact pairing (i.e., 1-1 correspondence) of the $n+2$ nodes of path $O(T)$ to the nodes $u_{n+1}, u_{n}, \ldots, u_1, u_{0}$ of the watermark graph $F[\pi^*]$;
\vspace*{0.05in}
\item[\textbf{4}.\,]
For each back-edge $(u_i,u_j)$ of the graph $F[\pi^*]$ (i.e., $u_j > u_i$), add the forward-edge $(v_j,v_i)$ in page-tree PT($T$) by adding in object ${\tt [v_j \ 0 \ obj]}$ an entry of type ${\tt /Key(v_i \ 0 \ R)}$; add in object ${\tt [v_{n+1} \ 0 \ obj]}$ an entry of type ${\tt /Key(v_0 \ 0 \ R)}$;
\vspace*{0.05in}
\item[\textbf{5}.\,]
Return the modified PDF document $T$, i.e., the watermarked document $T_w$;
\end{enumerate}

\noindent Let us briefly discuss the way we add forward-edge in the page-tree PT($T$); recall that, in Step~4 of the previous algorithm ${\tt Encode\_RPG.to.PDF}$ we add the forward-edge $(v_j,v_i)$ in page-tree PT($T$) by adding in object ${\tt [v_j \ 0 \ obj]}$ an entry of type ${\tt /Key(v_i \ 0 \ R)}$. The entry ${\tt /Key(v_i \ 0 \ R)}$ may be of various types; note that, ${\tt /Key(\cdot)}$ is used as parameter in our algorithm's description.

In our implementation, for the forward-edge $(v_j,v_i)$ such that the object ${\tt [v_j \ 0 \ obj]}$ is not the rood-node ${\tt root(pt)}$ of the page-tree PT($T$), we always chose the entry ${\tt /Key(v_i \ 0 \ R)}$ which we add in object ${\tt [v_j \ 0 \ obj]}$ to be of the same type of object ${\tt [v_i \ 0 \ obj]}$. In the case where $v_j = {\tt root(pt)}$, we chose the entry ${\tt /Key(v_i \ 0 \ R)}$ to be of type ${\tt /Kids(\cdot)}$.

For example, in Figure~\ref{fig:RPG-in-Page-tree} we have added forward-edges from object ${\tt [29 \ 0 \ obj]}$ to object ${\tt [3 \ 0 \ obj]}$, from object ${\tt [29 \ 0 \ obj]}$ to object ${\tt [24 \ 0 \ obj]}$, from object ${\tt [3 \ 0 \ obj]}$ to object ${\tt [13 \ 0 \ obj]}$, etc. Thus, in our implementation we have added in the root-node object ${\tt [29 \ 0 \ obj]}$ the entries ${\tt /Kids(3 \ 0 \ R)}$ and ${\tt /Kids(24 \ 0 \ R)}$, in object ${\tt [3 \ 0 \ obj]}$ the entry ${\tt /XObject(13 \ 0 \ R)}$, while in object ${\tt [13 \ 0 \ obj]}$ the entries ${\tt /ColorSpace(6 \ 0 \ R)}$ and ${\tt /R9(5 \ 0 \ R)}$.

\vspace*{0.2in}
\noindent {\bf Remark~3.1}.~Let $T$ be a PDF file and let PT($T$) be a page-tree of the document structure DS($T$). A node of the page-tree PT($T$) may contain several entries ${\tt /Key(\cdot)}$ of various types. We mention that, some types are required for the entries in specific nodes of PT($T$); for example, the required entries in the root-node ${\tt root(pt)}$ of the page-tree PT($T$) are the following four: ${\tt /Type(\cdot)}$, ${\tt /Parent(\cdot)}$, ${\tt /Kids(\cdot)}$, and ${\tt /Count(\cdot)}$.

\vspace*{0.2in}
\noindent {\bf Extraction.}~We next describe the corresponding extraction algorithm, which we call ${\tt Extract\_RPG.from.PDF}$;  it extracts the graph $F[\pi^*]$ from the PDF document $T_w$ watermarked by the embedding algorithm ${\tt Encode\_RPG.to.PDF}$. The algorithm works as follows:

\begin{enumerate}
\item[$\bullet$\,] Take first as input the PDF document $T_w$ watermarked by the embedding algorithm ${\tt Encode\_RPG.to.PDF}$, compute the document structure DS($T_w$) of $T_w$, and locate its page tree PT($T_w$); then, find in object $root(pt)$, where $root(pt)$ is the root of the tree PT($T_w$), the entry ${\tt /Kids(v_k \ 0 \ R)}$ s.t. $v_k$ is not a child of $root(pt)$, and set $v_{n+1}=root(pt)$ and $v_{0}=v_k$;
\vspace*{0.05in}
\item[$\bullet$\,] Compute the path $O(T)=(v_{n+1}, v_{n}, \ldots, v_1, v_{0})$ of PT($T_w$), from node $root(pt)$ to $v_{0}$, and assign an exact pairing (i.e., 1-1 correspondence) of the $n+2$ nodes of path $O(T)$ to the nodes $u_{n+1}, u_{n}, \ldots, u_1, u_{0}$ of a graph $F[\pi^*]$; initially, $E(F[\pi^*])=\emptyset$;
\vspace*{0.05in}
\item[$\bullet$\,] Add edges $(u_{i+1},u_{i})$ in $F[\pi^*]$ for $i=n, n-1, \ldots, 0$, and the edge $(u_{i},u_{j})$ iff $(v_{i},v_{j})$ is a forward edge in the page tree PT($T_w$);
\vspace*{0.05in}
\item[$\bullet$\,] Delete the edge $(u_{n+1},u_{0})$ from the graph $F[\pi^*]$;
\vspace*{0.05in}
\item[$\bullet$\,] Return the graph $F[\pi^*]$;
\end{enumerate}

\vspace*{0.03in}
\noindent It is easy to see that, by construction the returned graph $F[\pi^*]$ is a reducible permutation graph produced by either algorithm ${\tt Encode\_SiP.to.RPG}$-${\tt I}$ or algorithm ${\tt Encode\_SiP.to.RPG}$-${\tt II}$. Thus, $F[\pi^*]$ has the following property: the structure which results after deleting

\begin{itemize}
\item[(i)] all the forward edges $(u_{i+1},u_{i})$ of $F[\pi^*]$, $0 \leq i \leq n$, and

\item[(ii)] the node $u_{0}$
\end{itemize}

\noindent is either the tree~$T_d[\pi^*]$ or the tree~$T_{s}[\pi^*]$ produced during the execution of either the decoding algorithm ${\tt Decode\_RPG.to.SiP}$-${\tt I}$ or algorithm ${\tt Decode\_RPG.to.SiP}$-${\tt II}$, respectively (see, \cite{Chroni-PhD2014,CN13,CN12}). Thus, we can efficiently extract the self-inverting permutation $\pi^*$ embedded into a PDF document $T$ by algorithm ${\tt Encode\_RPG.to.PDF}$.

\section{Concluding Remarks}
\label{sec:Concluding-Remarks-PDF}

\noindent In this paper we presented embedded algorithms, along with their corresponding extraction algorithms, for watermarking PDF documents $T$ using three different representations of a self-inverting permutation $\pi^*$, namely 1D-representation, 2D-representation, and RPG-representation; note that, RPG-representation means the encoding of permutation $\pi^*$ as a reducible permutation graph $F^*[\pi^*]$.

The main features of our algorithms, i.e., the way they mark a PDF document $T$ or, equivalently, the way they embed a self-inverting permutation $\pi^*$ into document $T$, are summarized as follows:

\begin{itemize}
\item[$\circ$\,] In the first algorithm ${\tt Embed\_SiP.to.PDF}$-${\tt I}$ the marking of a PDF document $T$ is performed by increasing the distance (or, space) between two consecutive words in a paragraph of $T$.

\vspace*{0.05in}
\item[$\circ$\,] The main idea behind the second algorithm ${\tt Embed\_SiP.to.PDF}$-${\tt II}$ is based on the fact that $\pi^*$ has a 2D-representation and, since pages of a PDF documents $T$ are two dimensional objects, it can be efficiently used to mark specific positions on a page of $T$ resulting thus the watermarked PDF document $T_w$.

\vspace*{0.05in}
\item[$\circ$\,] The third graph-based embedding algorithm ${\tt Encode\_RPG.to.PDF}$ uses a deferent approach: it exploits the structure of a PDF document $T$ and embeds the graph $F[\pi^*]$ into $T$ by adding appropriate object-references in the document $T$, through the use of entries of type ${\tt /Kids(k \ 0 \ R)}$, so that the graph $F[\pi^*]$ can be easily constructed from the page tree PT($T_w$) of the resulting watermarked document $T_w$.
\end{itemize}

\noindent In light of our graph-based embedding algorithm ${\tt Encode\_RPG.to.PDF}$ it would be very interesting to investigate the possibility of altering other components of the document structure of a PDF file in order to embed the graph $F[\pi^*]$; we leave it as a direction for future work.

Moreover, an interesting open question is whether the embedding approaches and techniques used in this paper can help develop efficient encoding algorithms having ``better" properties with respect text attacks; we leave it as an open problem for future investigation.

\end{document}